# Abdominal aortic aneurysm wall stress: A 7-line code in MATLAB and a one-click software application

Mostafa Jamshidian[1*[0000-0002-5166-171X]], Saeideh Sekhavat[1[0009-0006-9280-2781]], Adam Wittek[1[0000-0001-9780-8361]], Donatien Le Liepvre[2], Florian Bernard[2], Ludovic Minvielle[2], Antoine Fondanèche[2], Karol Miller[1[0000-0002-6577-2082]]

[1] Intelligent Systems for Medicine Laboratory, The University of Western Australia, Perth, Western Australia, Australia
[2] Nurea, Bordeaux, France
mostafa.jamshidian@uwa.edu.au

**Abstract.** An abdominal aortic aneurysm (AAA) is a life-threatening condition characterized by the irreversible dilation of the lower aorta, usually detected incidentally during imaging for other health issues. Current clinical practice for managing AAA relies on a one-size-fits-all approach, based on the aneurysm's maximum diameter and growth rate, which can lead to underestimation or overestimation of AAA rupture risk. Patient-specific AAA wall stress, computed using biomechanical models derived from medical images without needing patient-specific material properties, has been widely investigated for developing individualized AAA rupture risk predictors. Therefore, AAA wall stress, determined reliably and quickly, has the potential to enhance patient-specific treatment plans. This paper presents a 7-line code, written in MATLAB using the Partial Differential Equation Toolbox, for AAA wall stress computations via finite element analysis. The code takes AAA wall geometry as input and outputs stress components over the AAA wall domain. Additionally, we present a one-click standalone software application for AAA wall stress computation, developed based on our 7-line code using MATLAB Compiler. After verification, we used our code to compute AAA wall stress in ten patients. Our analysis indicated that the 99th percentile of maximum principal stress across all patients ranged from 0.307 MPa to 0.466 MPa, with an average of 0.380 MPa and a standard deviation of 0.048 MPa. Moreover, for every case, the MATLAB simulation time was less than a minute on a laptop workstation.

## 1 Introduction

An abdominal aortic aneurysm (AAA) is an asymptomatic, permanent, and irreversible dilation of the lower aorta that is commonly diagnosed incidentally during imaging for other health conditions. If left untreated, AAA can expand to the point of rupture, leading to death most of the time [1, 2].

Currently, clinical practice for AAA management is a one-size-fits-all approach based on the aneurysm's maximum diameter and growth rate. Clinical intervention should be recommended when aneurysm diameters exceed 5.5 cm for men and 5 cm for women, or when the growth rate exceeds 1 cm annually [2]. As demonstrated by ruptured AAAs with diameters smaller than the critical diameter [3] and stable AAAs with diameters larger than the critical diameter [1, 2, 4], the maximum diameter criteria may underestimate or overestimate the rupture risk. Autopsy results indicate that approximately 13% of AAAs with a diameter of 5 cm or less ruptured, while 60% of AAAs larger than 5 cm in diameter did not rupture [5]. It is thus possible to improve lives and reduce healthcare costs by developing a patient-specific AAA rupture risk predictor.

Based on mechanics, an artery ruptures when the local wall stress exceeds the local wall strength. Hence, AAA wall stress, computed based on patient-specific biomechanical models extracted from medical images, has widely been investigated to develop a patient-specific AAA rupture risk predictor [6-15]. Significant progress in wall stress computation has been achieved through the invention of biomechanical models that provide wall tension without necessitating mechanical properties and thickness distribution of the wall [16]. AAA wall stress, determined reliably and quickly through non-invasive approaches, can potentially help with patient-specific treatment.

BioPARR (Biomechanics based Prediction of Aneurysm Rupture Risk) is a free software package for AAA stress analysis based on the finite element (FE) method [17, 18]. Fed with the patient’s blood pressure and the AAA and blood label maps extracted from the medical images, BioPARR conducts automatic robust computations based on the recent computational biomechanics approaches that obviate the requirement for patient-specific material properties [16]. As a truly patient-specific approach based solely on clinical data, BioPARR presents a clinically applicable standardized approach for the comparative evaluation of AAA.

While seamless at the time of release, BioPARR can be improved based on recent software advancements. BioPARR relies on multiple libraries and external software including the freely available 3D Slicer image computing platform [19, 20], the scientific visualization free software ParaView, the FE mesh generator free software Gmsh [21, 22], and the commercially-available Abaqus FE software [23]. BioPARR functions optimally with the recommended versions of these external software, but subsequent updates may compromise its robustness. Generally, the recommended older version of freeware may lack adequate support for new hardware and operating systems. Hence, BioPARR requires updates and maintenance, aligned with updates to the external software. Another limitation involves using expensive commercial software like Abaqus, which could potentially be substituted with freely available FE packages.

In this paper, we introduce a 7-line code and a one-click software application for AAA stress analysis written in MATLAB programming language. The remainder of the paper is organized as follows: In Section 2, we present the methods for AAA stress analysis in MATLAB and its verification against BioPARR stress analysis using Abaqus FE software [23]. We also describe the one-click software application developed using the 7-line code as the backend. In Section 3, we present the results for the mesh convergence analysis in MATLAB and verification of the MATLAB FE analysis. We also present the AAA stress analysis in ten patients. Finally, Section 4 is devoted to the conclusions and discussions.

# 2 Methods

## 2.1 Image data and AAA surface model

We employed anonymized contrast-enhanced 3D CTA images of ten patients diagnosed with AAA. These patients were recruited at Fiona Stanley Hospital in Western Australia and provided informed consent before participating in the study. The research was conducted in accordance with the Declaration of Helsinki, and the protocol received approval from the Human Research Ethics and Governance at South Metropolitan Health Service (HREC-SMHS) under approval code RGS3501, as well as from the Human Research Ethics Office at The University of Western Australia under approval code RA/4/20/5913.

As an example of the image data, Fig. 1a displays the cropped 3D CTA image of Patient 1's AAA. This image has dimensions of (112, 109, 74) voxels and voxel spacing of (0.627, 0.627, 1.000) mm along the (R, A, S) axes, as shown in Fig. 1a. In the patient coordinate system, the basis aligns with the anatomical axes of anterior-posterior, inferior-superior, and left-right. Specifically, the R, A, and S axes correspond to the left-right, posterior-anterior, and inferior-superior directions, respectively.

To create the patient's AAA surface model used in FE analysis, we used PRAEVAorta by NUREA (https://www.nurea-soft.com/) for AI-based automatic segmentation of AAA image in Fig. 1a, followed by automated post-processing with in-house MATLAB code [24], to automatically extract the AAA and blood label maps, as shown in Fig. 1b and Fig. 1c, respectively. We then used these label maps and the assumed wall thickness of 1.5 mm [25] as input for BioPARR to automatically extract the wall geometry and create the AAA surface model, as shown in Fig. 1d. It should be noted that the AAA surface model shown in Fig. 1d does not contain an intraluminal thrombus (ILT). We saved the surface model as an STL file, which was used as the input geometry for the MATLAB FE analysis.

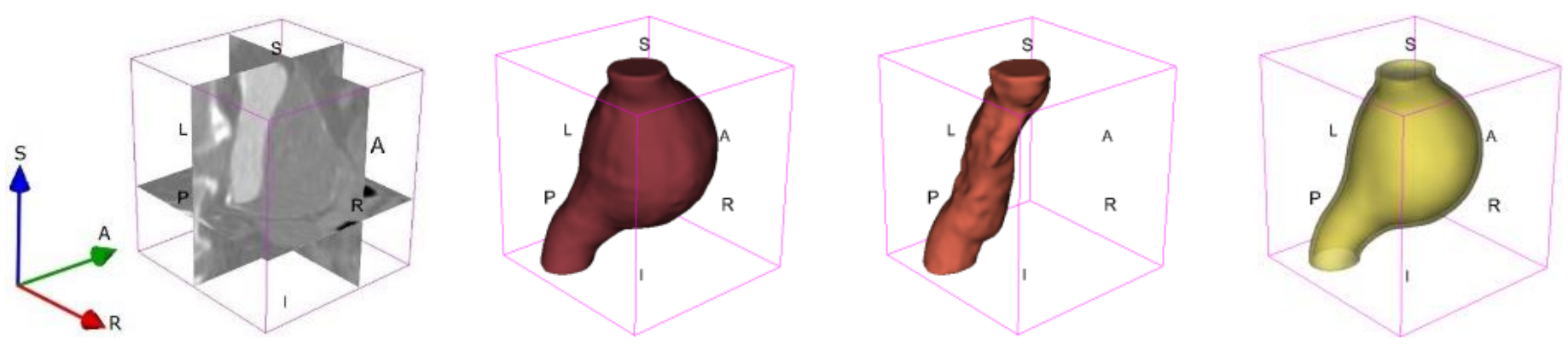

(a) (b) (c) (d)

**Fig. 1.** Patient 1's abdominal aortic aneurysm (AAA): (a) Cropped 3D CTA image, (b) AAA label map, (c) Blood label map, and (d) Surface model of AAA wall.

### 2.2 Stress analysis in MATLAB

The medical image data captures the geometry of a pressurized artery deformed by blood pressure. As a result, the artery geometries reconstructed from these images are the deformed configurations of the vessel structure. In biomechanics problems involving deformed geometries, material properties have minimal impact on wall stress [26, 27]. Consequently, we calculated the patient-specific AAA wall stress using a direct linear analysis without incorporating patient-specific material property data [16].

Starting from the AAA surface model shown in Fig. 1d, we conducted the stress analysis using MATLAB Partial Differential Equation (PDE) Toolbox. We performed AAA wall stress analysis in MATLAB via a 7-line code as follows:

```
structuralModel = createpde("structural","static-solid");

importGeometry(structuralModel,'WallSurface.stl');

structuralProper-
ties(structuralModel,"Cell",1,"YoungsModulus",100000,"PoissonsRatio",0.49);

structuralBC(structuralModel,"Face",[3,4],"Constraint","fixed");

structuralBoundaryLoad(structuralModel,"Face",1,"Pressure",13e-03);

generateMesh(structuralModel,"Hmax",1.0,'GeometricOrder','quadratic');

structuralresults = solve(structuralModel);
```

Line 1 creates a static structural model for solving a small-strain linear elasticity three-dimensional problem.

Line 2 creates a geometry object from the STL geometry file “WallSurface.stl” of the AAA wall surface model shown in Fig. 1d. The imported geometry object in MATLAB is shown in Fig. 2a. As shown in Fig. 2a, MATLAB automatically detects the cells and the faces of the imported geometry. For the AAA geometry, MATLAB detects one cell as C1 for the whole AAA wall geometry and four faces including the interior surface as F1, the exterior surface as F2, the bottom flat surface as F3, and the top flat surface as F4, as shown in Fig. 2a. We used these face labels to apply loads and boundary conditions.

Line 3 assigns isotropic linear elastic material properties for the structural model using Young modulus of 100 GPa and Poisson's ratio of 0.49. This high stiffness minimizes further deformation of the already deformed geometry of the AAA, with results remaining independent of Young modulus in stress recovery under a force-controlled scenario.

Line 4 applies the boundary conditions to the structural model by fixing the bottom (face F3) and the top (face F4) flat surfaces.

Line 5 applies a uniform pressure on the interior surface (face F1) of the aneurysm geometry using the patient-specific blood pressure of 13 kPa.

Line 6 meshes the geometry with quadratic tetrahedral elements using the target maximum element edge length of 1.0 mm, as shown in Fig. 2b.

Line 7 solves the FE model and returns the results including displacement, strain, and stress components at all nodes in the FE mesh. These primary results can be further processed to determine any desired quantities, such as maximum principal stress (MPS). As an example of the MATLAB FE analysis results, Fig. 2c shows the MPS contour plot on Patient 1's AAA wall.

For MATLAB FE analysis, we performed FE mesh convergence analysis using three different element sizes of 1.5, 1.0, and 0.5 mm, as shown in Fig. 3. The coarse FE mesh in Fig. 3a with FE size of 1.5 mm consists of 59962 nodes and 30364 elements. The FE mesh with medium mesh refinement using an FE size of 1.0 mm in Fig. 3b consists of 182506 nodes and 106675 elements. The fine FE mesh in Fig. 3c with an FE size of 0.5 mm consists of 1059351 nodes and 683966 elements.

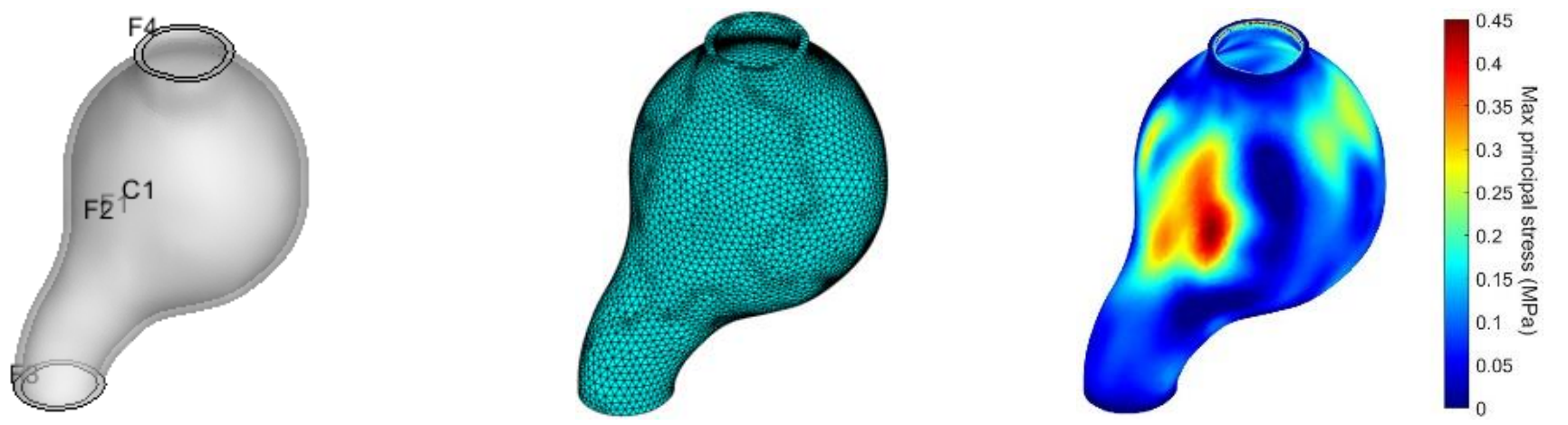

(a) (b) (c)

**Fig. 2.** Finite element (FE) stress analysis of abdominal aortic aneurysm (AAA) wall in MATLAB: (a) Surface model imported in MATLAB, (b) MATLAB FE Mesh, (c) MATLAB FE analysis results for the maximum principal stress (MPS) on AAA wall.

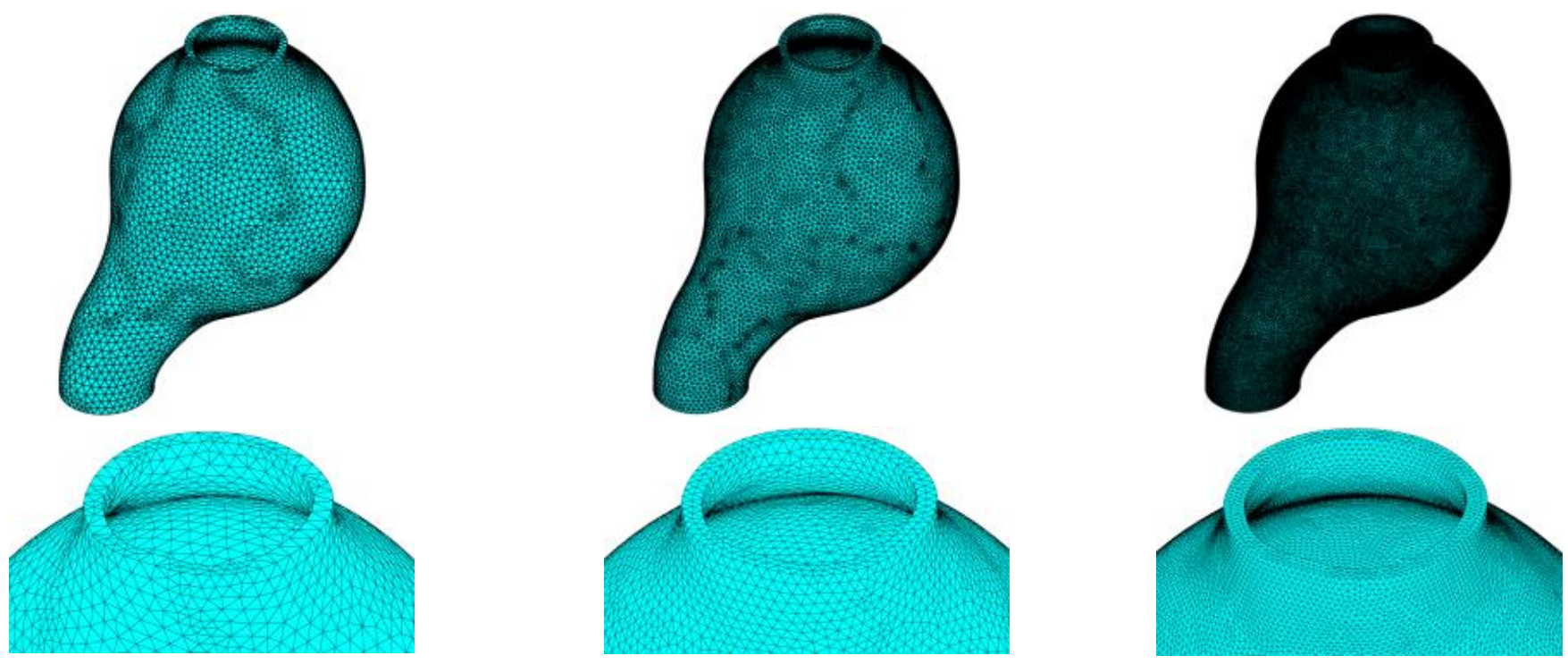

(a) (b) (c)

**Fig. 3.** Finite element (FE) mesh convergence Analysis in MATLAB using three different element sizes including (a) FE size 1.5 mm, (b) FE size 1.0 mm, (c) FE size 0.5 mm. Mesh refinement increases from left to right, with coarser meshes on the left and finer meshes on the right.

## 2.3 Verification against Abaqus

Alkhatib et al. [28] investigated the effects of FE formulation, FE shape functions, spatial integration scheme, and number of elements through the wall thickness, on the efficiency, robustness, and accuracy of AAA wall stress computations. They showed that a hexahedral mesh of quadratic elements with two element layers through wall thickness provides a sufficiently accurate AAA wall stress [28, 29].

We verified our MATLAB FE analysis against the previously verified Abaqus FE analysis [28]. Given the differences between the MATLAB mesh and the Abaqus mesh, we utilized stress percentile plots to compare two sets of stress results on the same domain (aneurysm wall) represented on different sets of discrete points [24, 30, 31]. In these plots, the horizontal axis represents the percentile rank, while the vertical axis shows the corresponding percentile value. Specifically, we verified the MATLAB FE analysis by comparing the percentile plots of the MPS on the exterior surface of AAA model in Fig. 1d.

In Section 3, we presented Abaqus FE analysis results for two FE meshes including a tetrahedral mesh automatically created by BioPARR and a hexahedral mesh manually created in HyperMesh [32]. Fig. 4a shows the Abaqus tetrahedral mesh using Abaqus C3D10 elements, which are 10-node quadratic tetrahedron elements. The Abaqus tetrahedral mesh comprises 520209 nodes and 298270 elements. Fig. 4b shows the Abaqus hexahedral mesh using Abaqus C3D20 elements that are 20-node quadratic hexahedron elements. The Abaqus hexahedral mesh comprises 49530 nodes and 8892 elements. As an example of the Abaqus FE analysis results, Fig. 4c shows the MPS contour plot on Patient 1's AAA wall.

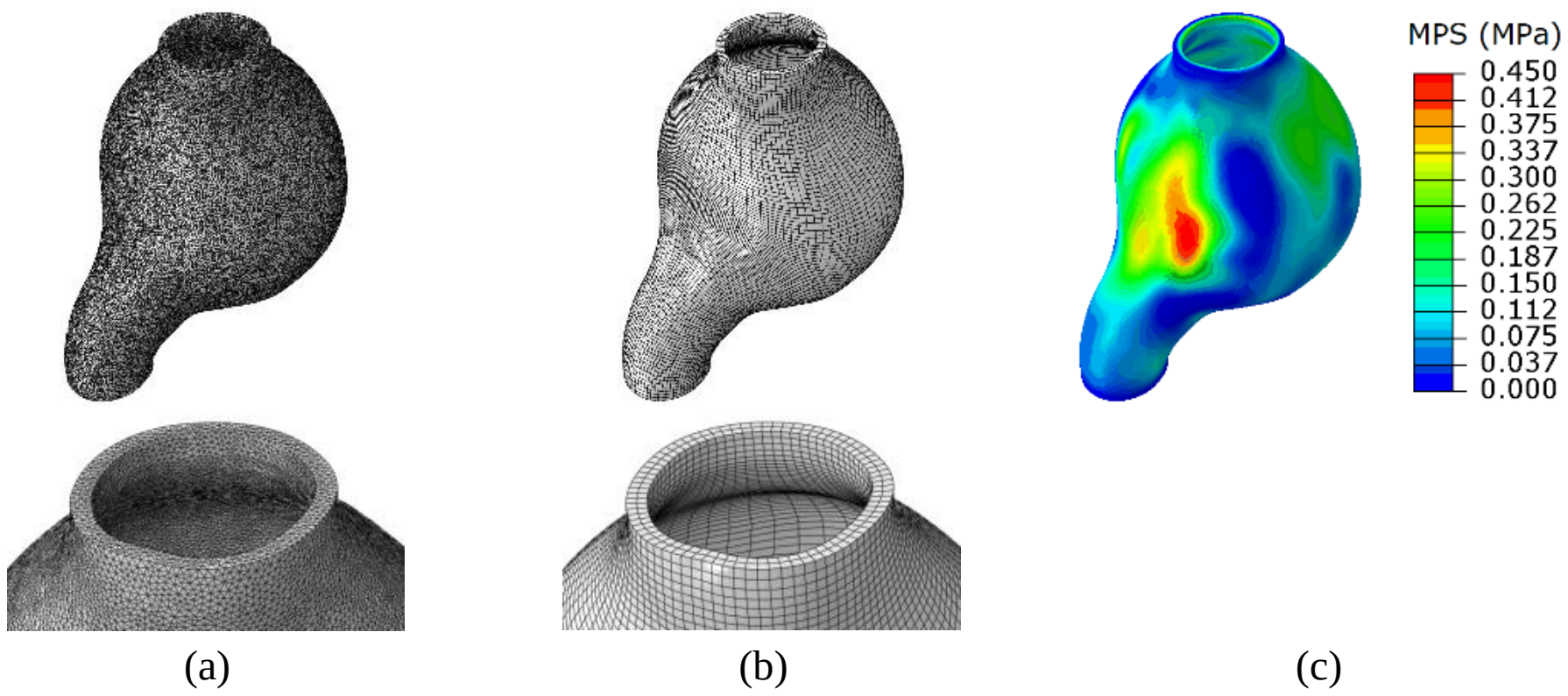

**Fig. 4.** Finite element (FE) stress analysis of abdominal aortic aneurysm (AAA) in Abaqus: (a) Tetrahedral FE mesh, (b) Hexahedral FE mesh, (c) Abaqus FE analysis results for the maximum principal stress (MPS) on AAA wall.

## 2.4 Software application

We developed a software application for AAA stress analysis, based on the 7-line code as the backend and using MATLAB Compiler™, which can be executed with just one click, as shown in Fig. 5. MATLAB Compiler™ enables sharing MATLAB programs as standalone applications. In addition, MATLAB Compiler™ encrypts the source MATLAB code files, ensuring that the source code remains hidden from the application users.

Our redistributable application for AAA stress analysis is a royalty-free standalone single executable file that is automatically installed on a typical system with no specific hardware requirements. During installation, it automatically fetches and installs MATLAB Runtime as a collection of shared libraries, MATLAB code, and other files that enable the execution of compiled and packaged MATLAB applications on systems without an installed version of MATLAB.

Fig. 5 shows the front end of our software application that provides a user-friendly window for the end user to input the required data. A single click on the application window initiates the backend computations and visualizes the AAA wall stress results. The backend executes our 7-line code for AAA wall stress computations that utilizes recent MATLAB capabilities, including the MATLAB PDE Toolbox.

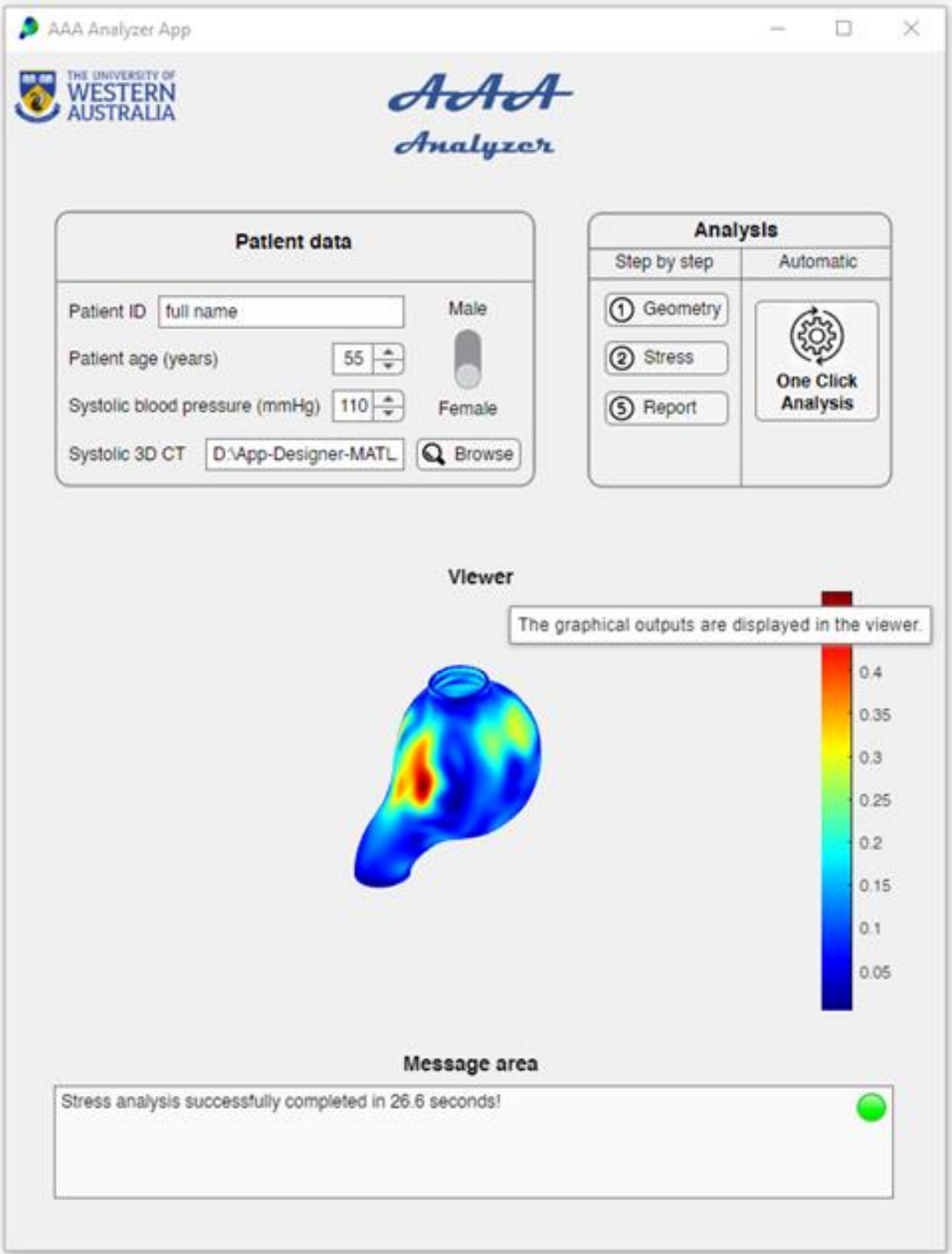

**Fig. 5.** Software application for one-click stress analysis of abdominal aortic aneurysm (AAA) developed in MATLAB.

## 3 Results

Fig. 6 presents the mesh convergence analysis results for the MATLAB FE analysis, using the FE mesh refinements depicted in Fig. 3. The variable of interest in the mesh convergence analysis was the MPS on the AAA exterior surface. In Fig. 6, the MPS percentile plots demonstrate convergence across different mesh refinements. The MPS contour plots, from left to right, correspond to increasingly finer meshes. Fig. 6 shows that the FE solution converges at an FE size of 1.0 mm.

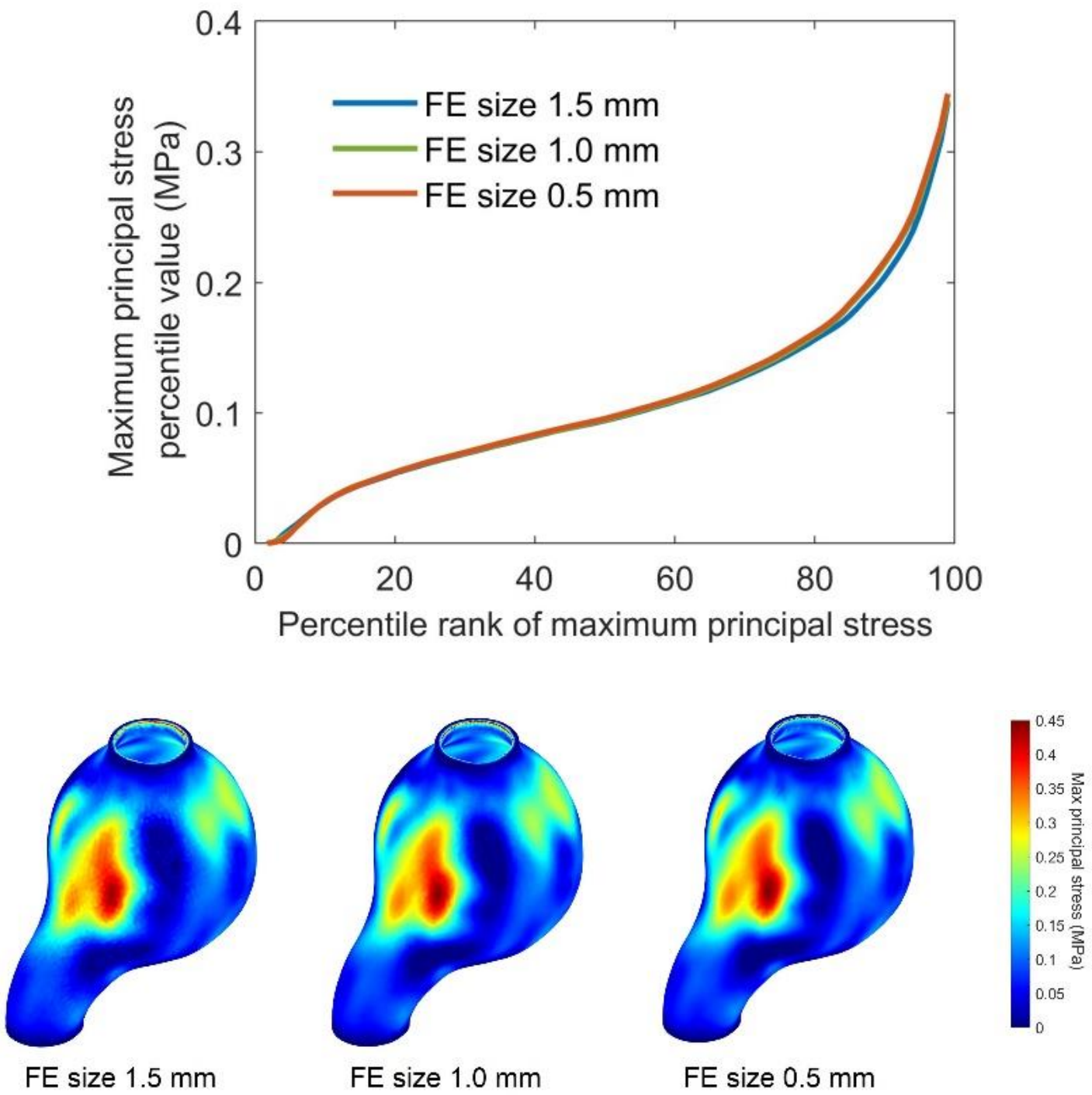

**Fig. 6.** Mesh convergence analysis for the MATLAB finite element (FE) analysis using the maximum principal stress (MPS) on the exterior surface of the abdominal aortic aneurysm (AAA) as the variable of interest. The MPS percentile plots show convergence across various mesh refinements. The MPS contour plots, from left to right, correspond to coarser to finer meshes.

Fig. 7 demonstrates the verification of MATLAB FE analysis against Abaqus FE analysis by comparing their MPS results on the exterior surface of Patient 1's AAA wall, shown in Fig. 1d. In addition to the MPS contour plots, Fig. 7 includes percentile plots of MPS, offering a detailed comparison between the MATLAB and Abaqus FE analysis results. The MATLAB FE analysis results in Fig. 7 are based on a tetrahedral mesh with an FE size of 1.0 mm. The Abaqus FE analysis results in Fig. 7 include the MPS for both the tetrahedral and hexahedral meshes, as shown in Fig. 4a and Fig. 4b, respectively.

Fig. 7 shows that the MATLAB FE analysis provides identical stress results to the Abaqus FE analysis using a tetrahedral mesh, across the entire percentile rank of maximum principal stress. The MPS from the Abaqus FE analysis using a hexahedral mesh is very close to that of the Abaqus FE analysis using a tetrahedral mesh.

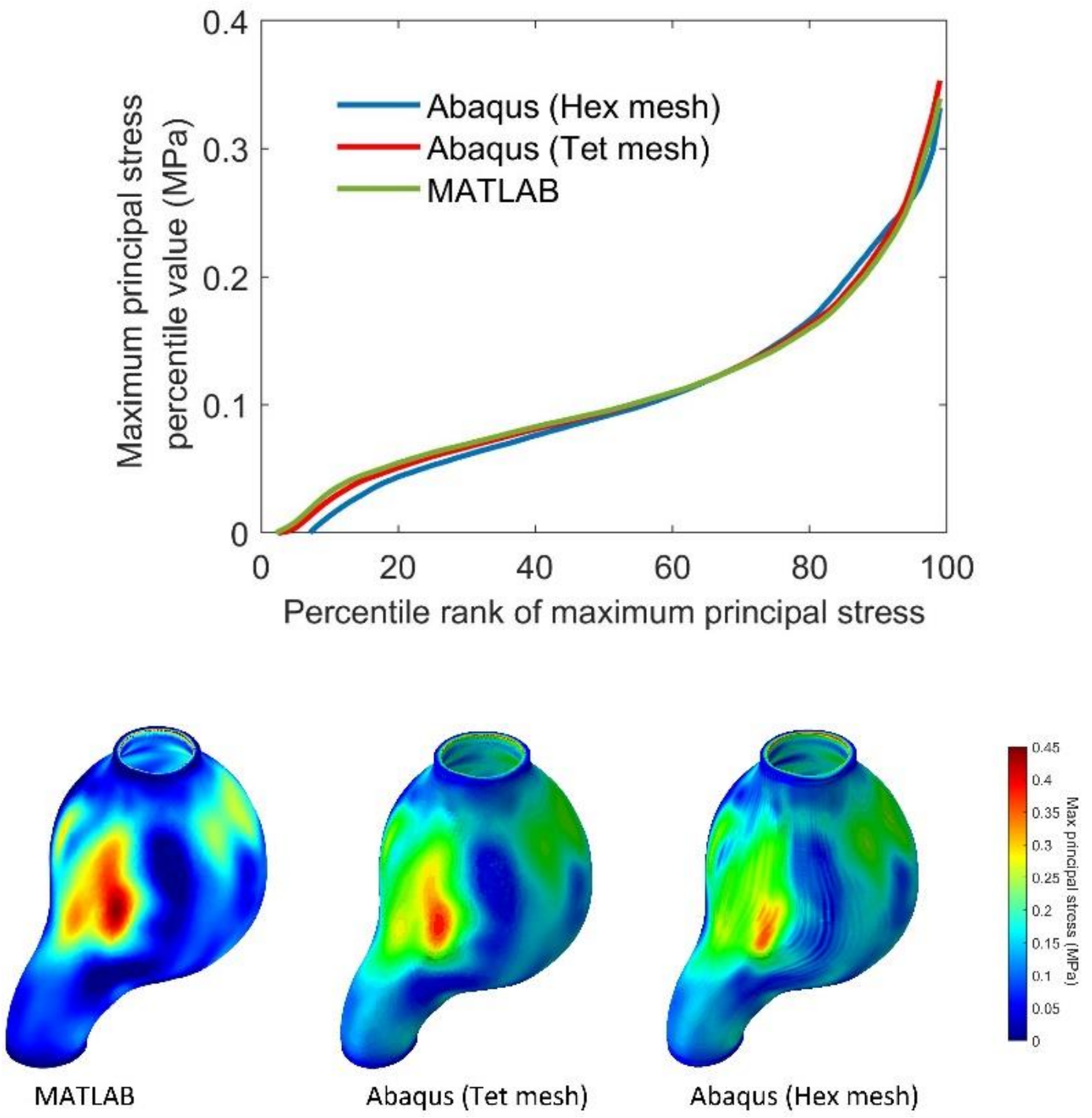

**Fig. 7.** Verification of MATLAB finite element (FE) analysis against Abaqus FE analysis by comparing the maximum principal stress (MPS) on the AAA exterior surface between MATLAB and Abaqus. The Abaqus results include the MPS for both the tetrahedral and hexahedral meshes. MPS percentile plots are presented alongside MPS contour plots to provide a detailed comparison between the different sets of results.

We employed our 7-line code to assess AAA wall stress in ten patients, as reported in Table 1 and Table 2. Table 1 presents the AAA wall geometry and the MPS contour plots for each patient, with the 99th percentile of MPS as the maximum contour limit. Table 2 provides the 99th percentile of MPS and the simulation time for the MATLAB FE analysis for each patient, conducted on a laptop workstation with a 12th Gen Intel® Core™ i9-12900H 2.50 GHz processor and 64.0 GB of RAM.

Table 2 indicates that the MPS 99th percentile among all patients ranged from 0.307 MPa to 0.466 MPa, with an average of 0.380 MPa and a standard deviation of 0.048 MPa, which is 13% of the average. This relatively low standard deviation-to-average ratio suggests consistency in AAA wall stress across all patients.

Table 2 also shows that the average MATLAB simulation time for AAA stress analyses is 38 seconds, with a maximum of 45 seconds and a minimum of 29 seconds. This demonstrates that the entire computations, from STL geometry to AAA wall stress, can be completed in under 1 minute.

**Table 1.** AAA wall stress analysis results for ten patients using MATLAB. For each patient, the AAA wall geometry, and the contour plots of maximum principal stress (MPS) are displayed, with the 99th percentile of MPS used as the maximum contour limit.

| Patient number | AAA wall geometry | Maximum principal stress (MPa) | | |
|---|---|---|---|---|
| 1 | 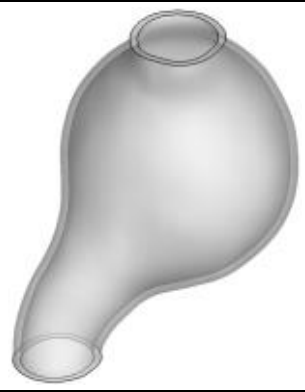 | 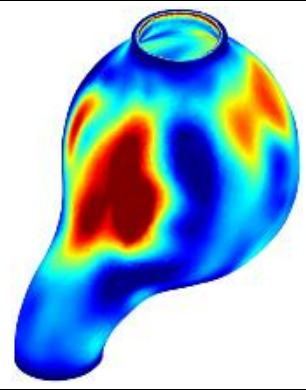 | 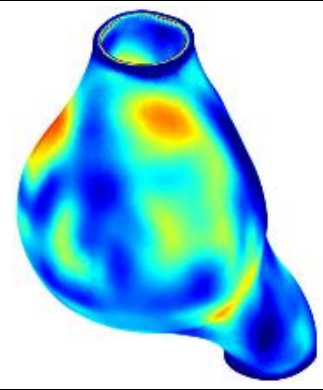 | 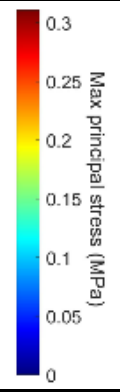 |
| 2 | 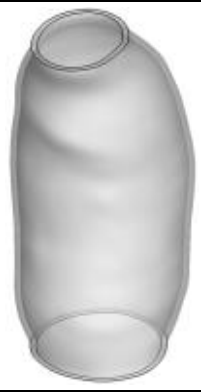 | 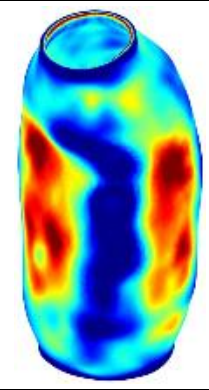 | 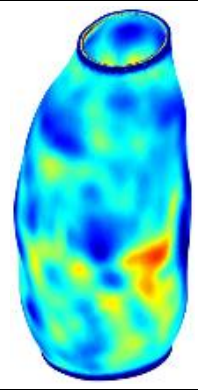 | 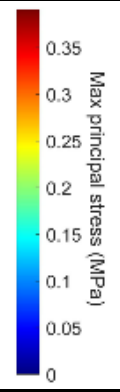 |
| 3 | 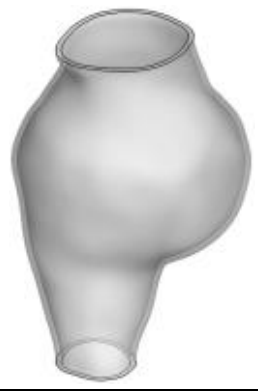 | 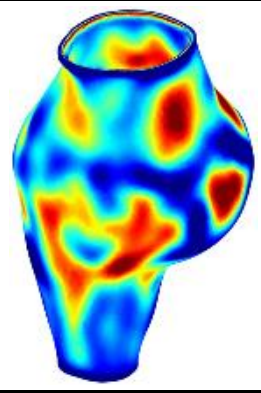 | 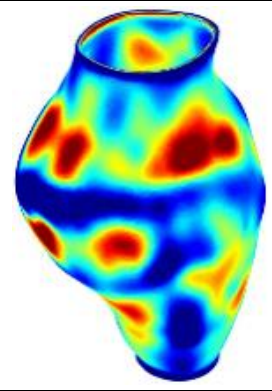 | 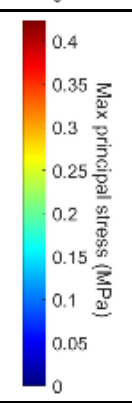 |
| 4 | 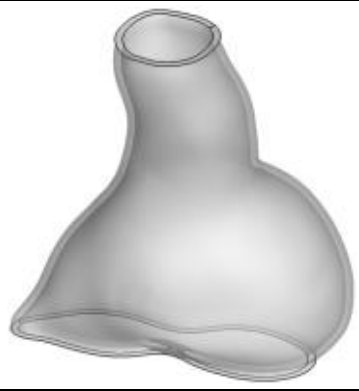 | 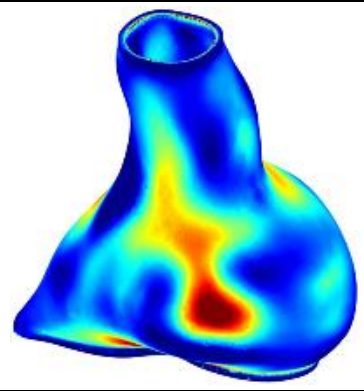 | 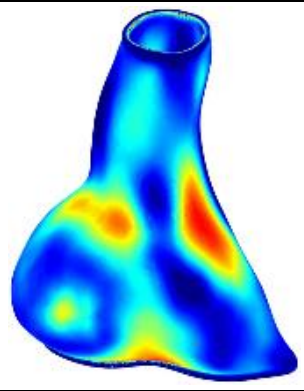 | 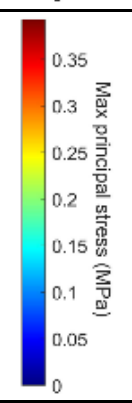 |
| 5 | 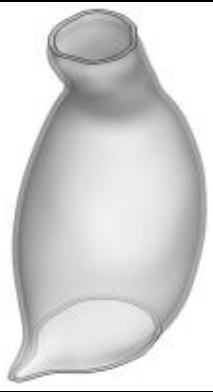 | 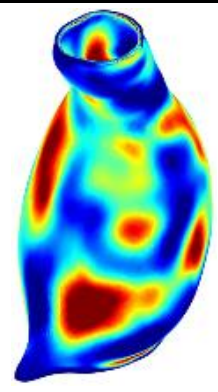 | 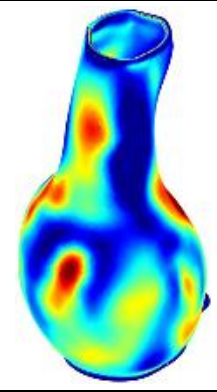 | 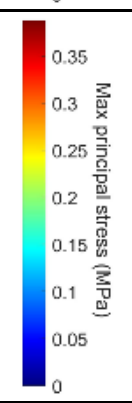 |
| 6 | 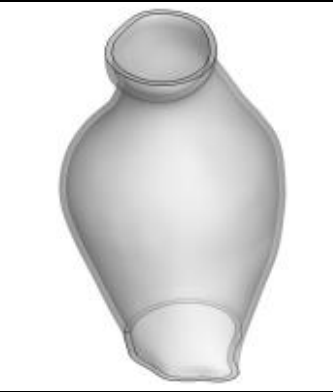 | 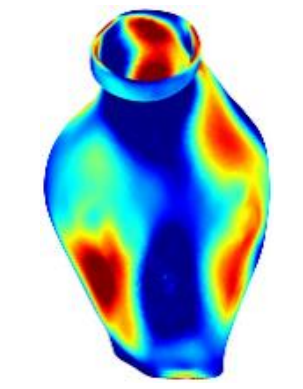 | 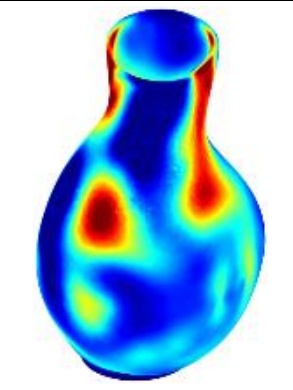 | 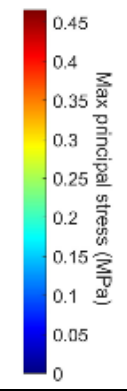 0.45 0.4 0.35 0.3 0.25 0.2 0.15 0.1 0.05 0 Max principal stress (MPa) |
| 7 | 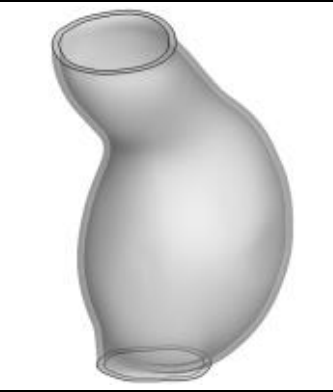 | 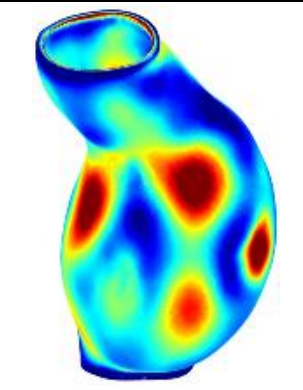 | 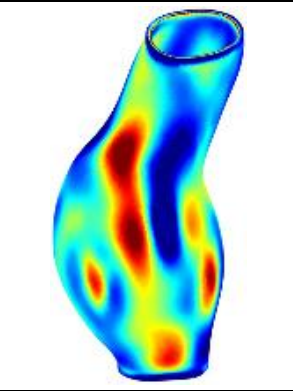 | 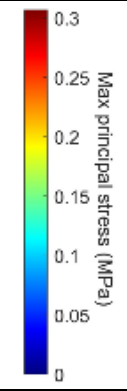 0.3 0.25 0.2 0.15 0.1 0.05 0 Max principal stress (MPa) |
| 8 | 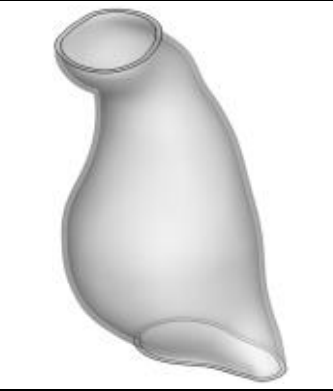 | 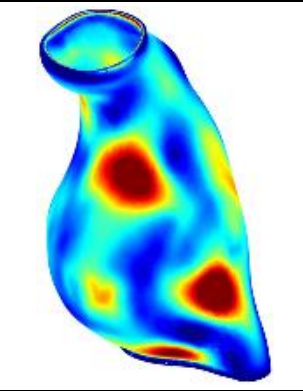 | 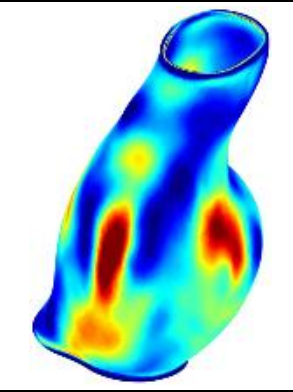 | 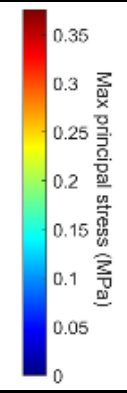 0.35 0.3 0.25 0.2 0.15 0.1 0.05 0 Max principal stress (MPa) |
| 9 | 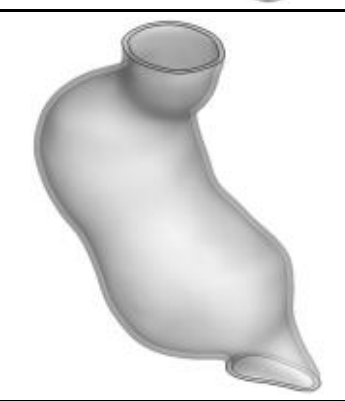 | 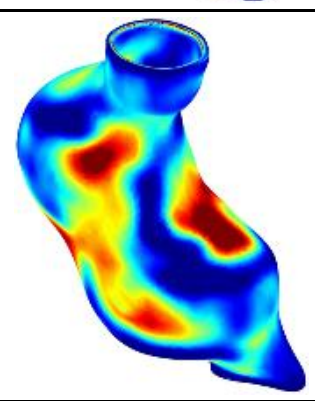 | 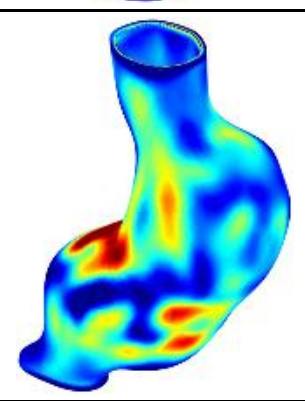 | 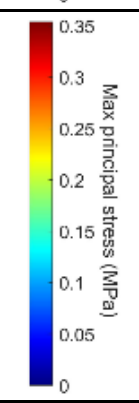 0.35 0.3 0.25 0.2 0.15 0.1 0.05 0 Max principal stress (MPa) |
| 10 | 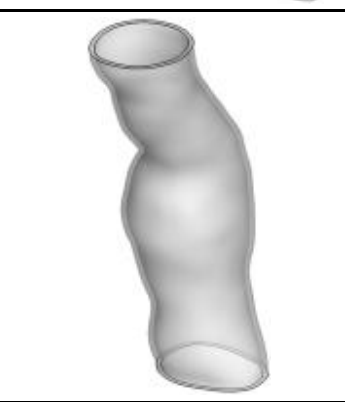 | 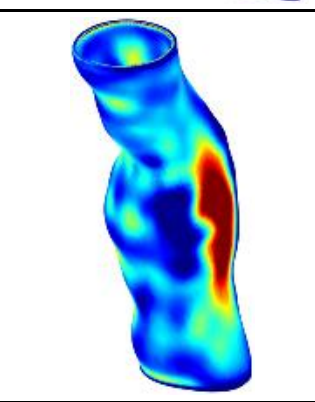 | 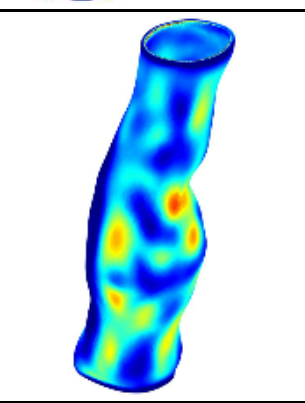 | 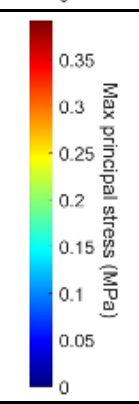 0.35 0.3 0.25 0.2 0.15 0.1 0.05 0 Max principal stress (MPa) |

**Table 2.** AAA wall stress analysis results for ten patients using MATLAB. For each patient, the 99th percentile of maximum principal stress (MPS) on the AAA wall and the finite element (FE) simulation time in MATLAB are reported.

| Patient number | MPS 99th percentile (MPa) | Simulation time (s) |
|---|---|---|
| 1 | 0.311 | 36 |
| 2 | 0.391 | 41 |
| 3 | 0.424 | 43 |
| 4 | 0.393 | 29 |
| 5 | 0.388 | 45 |
| 6 | 0.466 | 32 |
| 7 | 0.307 | 35 |
| 8 | 0.376 | 42 |
| 9 | 0.354 | 40 |
| 10 | 0.392 | 37 |
| Minimum | 0.307 | 29 |
| Maximum | 0.466 | 45 |
| Average | 0.380 | 38 |
| Standard deviation | 0.048 | 5 |

## 4 Conclusions and Discussions

We introduced a 7-line MATLAB code utilizing the Partial Differential Equation (PDE) Toolbox for computing AAA wall stress through finite element (FE) analysis. Additionally, using MATLAB Compiler, we developed a standalone one-click software application for AAA wall stress computation, with our 7-line code running in the backend. The input to our 7-line code is the AAA wall surface model derived from medical images, and its output is the stress components over the AAA wall. Our code utilizes a stress computation method, previously implemented in the free software package BioPARR (Biomechanics based Prediction of Aneurysm Rupture Risk), that eliminates the need for patient-specific material properties.

We performed a mesh convergence analysis for our MATLAB FE analysis and demonstrated that a tetrahedral mesh with quadratic elements of size 1.0 mm provides sufficiently accurate results for the maximum principal stress (MPS). We verified our MATLAB FE analysis against previously validated Abaqus FE analysis.

We applied our 7-line code to compute AAA wall stress in ten patients. The FE analysis results indicated that the 99th percentile of MPS across all patients ranged from 0.307 MPa to 0.466 MPa, with an average of 0.380 MPa and a standard deviation of 0.048 MPa. Furthermore, the entire MATLAB simulation time for each case was under a minute using a laptop workstation.

We assumed linear elastic, isotropic material properties for the AAA wall, constant wall thickness, and did not include residual stress or intraluminal thrombus (ILT) in our analysis. However, these assumptions and simplifications do not compromise the validity of our results. We have previously shown that, from the perspective of distin-guishing between AAAs that would rupture and those that would remain intact, linear elastic analysis using isotropic material properties yields practically the same results as more complicated non-linear models [30].

Low image resolution and poor soft tissue contrast impair the accuracy and reliability of aortic wall thickness measurements from CT images. Consequently, in this and many other studies of AAA biomechanics, constant wall thickness has been used [3, 10, 33]. BioPARR [17, 18] utilizes a very efficient method for computing residual stress based on the uniform stress hypothesis by Fung [34]. This method could have been incorporated into the approach and computer code presented in this study. However, including residual stress makes the distribution of stress within the AAA wall more uniform, which would be inconsistent with the purpose of verifying AAA stress determined using our fully automated approach against the results obtained using procedures available in established general-purpose finite element codes (we used Abaqus). Therefore, in this study, we did not account for the effects of residual stress.

The analysis of the effects of intraluminal thrombus is beyond the scope of this study. However, the automated approach for creating patient-specific AAA surface models from CT images, described in Section 2.1 (Image Data and AAA Surface Model), has been previously successfully applied to AAAs with ILT [24].

One major limitation of our method is the need for wall strength to determine stress-based rupture risk indicators [35, 36]. Since it is clinically impractical to extract in-vivo wall strength, population-based statistics are used to estimate wall strength, which do not accurately reflect individual patient conditions [37]. Recent research suggests that the dimensionless ratio of wall stress to wall strength at baseline is a predictor of AAA rupture or repair, improving risk stratification for AAA events compared to solely using the maximum diameter [35, 36].

However, the fact that the material properties of the wall vary among patients and locally within an individual's aortic wall significantly affects rupture risk estimation [38, 39]. Obviously, direct non-invasive measurement of in-vivo local patient-specific wall strength (or degradation) in a living individual is impossible, but it may be indirectly estimated by supplementing the wall stress map with the wall strain map [40]. Towards a more predictive rupture risk estimation, there have been attempts at non-invasive, in-vivo measurement of AAA wall strain using sequential images at different phases of the cardiac cycle [40-43].

Despite these limitations, our 7-line MATLAB code and standalone software application offer the clear advantage of fast and reliable AAA wall stress computation.

## Acknowledgements

This work was supported by the Australian National Health and Medical Research Council NHMRC Ideas grant No. APP2001689. Contributions of Christopher Wood and Jane Polce, radiology technicians at the Medical Imaging Department, Fiona Stanley Hospital (Murdoch, Western Australia) to patient image acquisition at Fiona Stanley Hospital, and Dr Farah Alkhatib’s (The University of Western Australia) assistance in patient recruitment and acquisition of these images are gratefully acknowledged.